\documentclass[twocolumn,showpacs,preprintnumbers,amsmath,amssymb]{revtex4}

\usepackage{graphicx}
\usepackage{dcolumn}
\usepackage{bm}

\begin{document}

\hyphenation{Es-ta-dual}

\preprint{BA-02-20}

\title{Hybrid simulations of extensive air showers}

\author{Jaime Alvarez-Mu\~niz}
\email{alvarez@bartol.udel.edu}
\author{Ralph Engel}
\author{T.K. Gaisser}
\author{Jeferson A. Ortiz}
\altaffiliation[Also at ]{Instituto de F\'{\i}sica ``Gleb Wataghin'',
Universidade Estadual de Campinas 13083-970 Campinas-SP, Brazil.
}
\author{Todor Stanev}
\altaffiliation[Also at ]{Laboratoire de Physique 
Corpusculaire et Cosmologie\\Coll\`ege de France, Paris, France.
}
\affiliation{Bartol Research Institute,\\
University of Delaware, Newark, Delaware 19716, U.S.A.\\
}


\begin{abstract}
 We present a fast one dimensional hybrid method to efficiently
 simulate extensive air showers up to the highest observed
 energies. Based on precalculated pion showers and a bootstrap
 technique, our method predicts the average shower profile,
 the number of muons at detector level above several energy
 thresholds as well as the fluctuations of the electromagnetic 
 and hadronic components of the shower. We study the main 
 characteristics of proton-induced
 air showers up to ultra-high energy, comparing the predictions
 of three different hadronic interaction models: SIBYLL 1.7,
 SIBYLL 2.1 and QGSjet98.  The influence of the hadronic interaction
 models on the shower evolution, in particular the elongation rate,
 is discussed and the applicability of analytical approximations 
 is investigated.
\end{abstract}



\maketitle

\section{\label{introduction}Introduction}

 Extensive air showers (EAS) generated by cosmic rays in the
 Earth's atmosphere are the only way to study cosmic rays of energies
 above $10^{15}$ eV. At lower energies the cosmic ray spectrum
 and composition are studied in experiments that measure directly the
 charge and energy of the primary particle.
 The analysis of air shower data relies on simulations that use
 the current knowledge of hadronic interactions to predict the
 observable shower parameters.
 With increasing cosmic ray energy,
 this task becomes more difficult as the gap between 
the shower energy and the energy range studied in accelerator
 experiments increases and the hadronic
 interaction properties have to be extrapolated over a wide range.
 The difficulties are also related to the fact that particles
 produced in the forward region of the interaction are not
 registered in collider experiments, while they are responsible
 for most of the shower characteristics.
 Last but not least, the atmospheric 
 targets are light nuclei which have not been studied
 in collider experiments.

 Air shower experiments are either ground arrays of detectors that
 trigger in coincidence when the shower passes through them, or
 optical detectors that observe the longitudinal development of EAS.
 Both types of instruments are sometimes supplemented by shielded or 
 underground detectors that observe the muon component of the showers.
 The most commonly observed EAS parameters are the number of charged
 particles at ground level for the shower arrays, or at
 shower maximum ($S_{\rm max}$) for the optical detectors; the
 depth of shower maximum ($X_{\rm max}$) itself, and the number of 
 muons ($N_\mu$) above different energy thresholds. The combination
 of these and occasionally additional shower features, calculated
 in simulations with a particular hadronic model, is used as the basis
 for the determination of the energy and mass of the primary particle.
Reviews of air shower experiments 
and observed features are given,
for example, in \cite{NaganoWatson00,Kampert:2000ne}.
 
At the end of the cosmic ray spectrum, at energies above
10$^{19}$ eV, air shower simulation becomes a very difficult
problem technically.
The number of charged particles that have to be followed in the
Monte Carlo scheme is proportional to the shower energy. 
For example, highest energy cosmic ray showers~\cite{Bird95,Takeda98}
can have more than $10^{11}$ charged particles at $X_{\rm max}$. As 
a consequence the direct
simulation of the shower following each individual particle becomes
 practically impossible, especially when a large number of showers
has to be simulated.  
  
The widely used solution to the problem of having to deal with an
excessively large number of
 shower particles is the simulation of EAS using the thinning
technique \cite{Hillas97a}. This method 
is extremely useful to estimate detectable signals 
and to compute average values of the observables
\cite{Shirasaki01,Capdevielle00}.   
 The thinning procedure follows only a subset of the 
 shower particles below a certain energy threshold, 
 assigning weights to them so that the 
 average number of particles at the ground is correctly
 reproduced.
Due to this, artificial fluctuations are introduced  
even when small energy thresholds are used.
Various methods of reducing
artificial fluctuations have been proposed recently (e.g.
\cite{Kobal01,Risse01}) optimizing the compromise between
time-consuming simulations and fluctuation-enhancing thinning. 

In this work we present a hybrid method of simulating 
the longitudinal profile of extensive air showers. 
It is a fast, one dimensional calculation which provides
predictions for the total number of charged particles
and muons along the shower axis.
The method allows the collection of sufficiently
high Monte Carlo statistics without losing information 
about shower fluctuations.

In general, hybrid calculations are based on the idea to follow the 
development of air showers in detail above
a certain energy threshold and to replace subthreshold  
particles by a simplified and efficient
approximation of the subshowers initiated by them.
Many hybrid calculations use the Monte Carlo-generated high-energy
secondary particles of the first few interactions of a cosmic ray in the
atmosphere as initial distribution, and then calculate the particle densities
observed at detector level by solving the corresponding transport equations
(see, for example, 
\cite{Dedenko68a,Kalmykov01,Kirillov01a,Kirillov01b,Farrar02}).

Here we follow the approach of 
Gaisser et al. \cite{Gaisser97} and
treat the subthreshold particles with
a library of shower profiles based on presimulated
pion-initiated showers. This idea can be combined in a 
bootstrap procedure~\cite{Stanev94} to extend the shower 
library to high energy. 
The novelty of this work is that 
we extend the method of \cite{Stanev94,Gaisser97} by
accounting for fluctuations in the subshowers generated with the 
shower library, and also calculate the number of muons 
at detector level above several energy thresholds.

Showers simulated in this way can be used as input to
simulations for experiments measuring the longitudinal
development of the shower such as HiRes~\cite{Abu-Zayyad00},
the fluorescence detector
of the Pierre Auger Observatory~\cite{Cronin95} and future 
experiments such as EUSO~\cite{Catalano01}, 
OWL/AirWatch~\cite{Streitmatter97} and the
Telescope Array~\cite{Sasaki01}. Besides this, as will become
 clear later, hybrid simulations are
very helpful for comparing shower parameters predicted by different
hadronic interaction models and to aid the interpretation of the
experimental results in this way.

This paper is structured as follows: In Section~\ref{HybridMethod}
we describe the hybrid method and the parametrizations
of the presimulated showers. We demonstrate the self-consistency of the 
method by comparing showers simulated directly with predictions from the
hybrid calculation.
In Section~\ref{ResultsProtons} we apply the hybrid method 
to proton induced showers at fixed energy.  
We give the average values and distributions of $X_{\mathrm {max}}$, 
$S_{\mathrm {max}}$, and $N_{\mathrm {\mu}}$ 
obtained for different hadronic models and discuss how the
differences are related to the simulation of high-energy 
multiparticle production. In addition the elongation rate
theorem \cite{Linsley77a,Linsley80a} is discussed in terms of the
different hadronic interaction models and their influence on the
position of the shower maximum.
Where available, we compare our predictions to calculations
performed with the CORSIKA code~\cite{Heck98} which uses
the thinning approach.
 Section~\ref{Summary}
 summarizes our results and concludes the paper. 

\section{\label{HybridMethod}The Hybrid Method}

The hybrid method used in this work consists of calculating shower 
observables
 by a direct simulation of the initial part of the shower,
 tracking all particles of energy $> f E$, where $E$ is
 the primary energy and $f$ is an appropriate fraction of it
 (in the following we use $f$ = 0.01). Then 
 presimulated  showers for all subthreshold  particles are 
 superimposed after
 their first interaction point is simulated. 
 The subshowers are described with
 parametrizations  that give the correct average
 behavior and at the same time describe the fluctuations 
 in shower development.
 The method is extended recursively to higher
 energies by bootstrapping. The results obtained   
 at any primary energy 
 can then be used for the simulation of showers at 
 higher energy.

 It is well known that the fluctuations in shower properties 
 are dominated by fluctuations in the earliest and most energetic
 part of the cascade. We however parametrize both the average
 behavior and the shower fluctuations starting at 10 GeV.
 In this way we can use the hybrid method at relatively low energy of
 100 -- 1000 TeV, where the results can be compared to those
 of direct (fully simulated) shower calculations. 

 We build a library of presimulated showers by injecting pions
 of fixed energy $E_\pi$, at fixed zenith angle $\theta$ and 
 depth $X$ measured along the shower axis. 
 The atmospheric density
 adopted here corresponds to Shibata's fit of the 
 US Standard Atmosphere~\cite{Forsythe69,Shibata81}, very similar
 to Linsley's parametrization. We limit
 the injection zenith angles to  
 $\theta < 45^\circ$ since mainly showers in this angular range 
 have been used for studies of the cosmic ray energy spectrum at the
 highest energies. 

Nucleon initiated showers are not presimulated.  
 Nucleons are followed explicitly in the
 Monte Carlo down to the energy threshold for particle production.
 A subshower initiated by a kaon is assumed to be similar to one
 initiated by a pion of the same energy but with a different
first interaction point, which is sampled from the corresponding
interaction length distribution.
 This approximation is 
 not expected to affect significantly our final results,  
 the main reasons being the similarity between 
 pion and kaon induced showers at high energy combined with the fact that
 the main contribution  
 to shower development in this method comes from the highest 
 energy particles that enter the parametrizations. 
Unstable particles, including $\pi^0$,
$\eta$, $\Lambda$, $\Sigma$ and $\Omega$ 
are allowed to interact or decay in the code.
The interaction of these particles becomes important at 
the highest energies
and accounting for them can influence the average values of
some observables. 

 Photon and electron/positron induced cascades are treated with a full
 screening electromagnetic Monte Carlo in combination with a modified
 Greisen parametrization. The electromagnetic branch of the Monte Carlo
 includes photoproduction of hadrons. For energies above
 1 EeV, the Landau-Pomeranchuk-Migdal (LPM) 
 effect~\cite{Landau1a,Landau2a,Migdal1a,Klein97} is taken into account
using an implementation by Vankov~\cite{Vankov}. 
 The influence of the geomagnetic field
 on the cascade development~\cite{Stanev97b} is neglected.

\begin{figure}
\centerline{ 
\includegraphics[width=8.5cm]{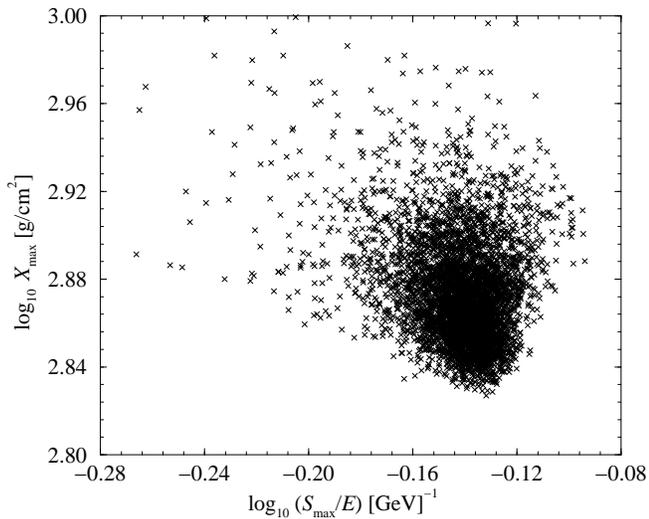} 
}
\caption{The correlation
 between $\lg X_{\mathrm {max}}$
and  $\lg S_{\mathrm {max}}$ for 
5,000 pion induced showers at primary energy $3\times10^{18}$ eV 
initiated at $X_0=5$ g/cm$^{2}$ and zenith angle
$\theta=45^\circ$.
}
\label{correlation}
\end{figure}
 We have simulated primary pions of energies between 
 10 GeV and 3 EeV with a step in energy of half a decade, interacting at
 fixed atmospheric depths $X_0=$ 5, 50, 100, 200, 500 and
 800 ${\mathrm {g/cm^2}}$.  
 For each pion energy, injection zenith angle and depth 
 (i.e. a single entry in the library) we simulate 10,000 showers
 (5,000 at high energy) and record $X_{\mathrm {max}}$,
 $S_{\mathrm {max}}$, the longitudinal shower profile,
 and the number of muons above the threshold
 energies of 0.3, 1, 3, 10 and 30 GeV both at sea level and at 
 a depth of $400~{\rm g/cm^2}$ above sea level measured 
along the shower axis. 
 These values are used to produce distributions of showers in 
 $X_{\mathrm {max}}$ and $S_{\mathrm {max}}$, the correlations
 between them and distributions of the number of muons at 
 sea level. The whole procedure of generating a library
 has to be carried out for each of the interaction models
 we adopt in this work (see section~\ref{Models}).

 Fig.~\ref{correlation} shows an example of the correlation
 between  $X_{\mathrm {max}}$
 and $S_{\mathrm {max}}$. The plot contains 5,000 simulated
 pion showers of energy $3.16\times 10^{18}$ eV initiated
 at atmospheric depth of $X_0$=5 g/cm$^2$ and zenith angle
 $\theta = 45^\circ$. Correlations similar to these are produced
 for each entry in the library.
 Their correct representation is
 crucial for the successful modeling of shower fluctuations.

 Although it is unlikely to produce a high energy pion deep in 
 the atmosphere, we also calculate their interactions at 
 depths as large as 500 and 800 ${\rm g/cm^2}$ 
 to obtain an accurate description of the muon numbers at sea level
 and a better description of the late developing electromagnetic
 showers. For showers initiated after $500~{\rm g/cm^{2}}$ the
 atmosphere has been artificially extended beyond ground level.
The distributions of muons are easily extended
to other depths (corresponding to the observation level of different
experiments) by extrapolation. For this task we use the slope of
the muon longitudinal profile between sea level and
a slant depth of $400~{\rm g/cm^2}$ above sea level.

 The longitudinal development of 
subthreshold meson induced showers is parametrized
using a slightly modified version of the well-known Gaisser-Hillas
function that gives the number of charged particles at atmospheric 
 depth $X$,~\cite{Gaisserbook}:
\vspace{-0.1cm}
\begin{eqnarray}
S_{\rm GH}(X)=S_{\rm max}~
\left( {X-X_0 \over X_{\rm max}-X_0} \right)^
{(X_{\rm max}-X_0)/\lambda(X)}\nonumber\\
\times~{\rm exp}
\left[-{(X-X_{\rm max})\over\lambda(X)}\right].
\label{eq:GH}
\end{eqnarray}
Here $\lambda(X)=\lambda_0+bX+cX^2$ where $\lambda_0$,
$b$ and $c$ are treated as free parameters. $X_0$ is the depth at
 which the first interaction occurs. 
The parameters $b$ and $c$ are assumed to be the same for 
all showers initiated at a given
depth, angle, and energy. They are determined 
by fitting the mean shower profile of the parametrized showers to 
that obtained from the simulated shower profiles.

The innovative approach
 of our method is that instead of using the
 average values of $X_{\mathrm{max}}$ and $S_{\mathrm{max}}$
 to generate subthreshold
 meson showers of a certain energy, we sample their values (as 
well as the number of muons) from
 their corresponding presimulated distributions, taking into
 account the correlation between them (Fig.~\ref{correlation}). 
 This procedure accounts for the fluctuations in the subshower
 development. 
A technical remark is that we sample the observables directly
from their precalculated histograms, i.e. we do not assume 
any functional form for the distribution. In this way our
code is very flexible -- it allows the study of hadronic 
models that predict distributions of observables 
not easily fitted by  analytical functions.  

We sample meson subshowers at a zenith angle, depth and/or 
primary energy different from those we have presimulated by 
interpolating between the relevant parameters of the 
shower development ($X_{\mathrm{max}}, S_{\mathrm{max}}, X_0, b, c, N_\mu$), 
corresponding to presimulated entries  
in the library which are adjacent in angle, energy, and depth
to the subshower we want to describe.  

\begin{figure}[h]
\centerline{
\includegraphics[width=8.5cm]{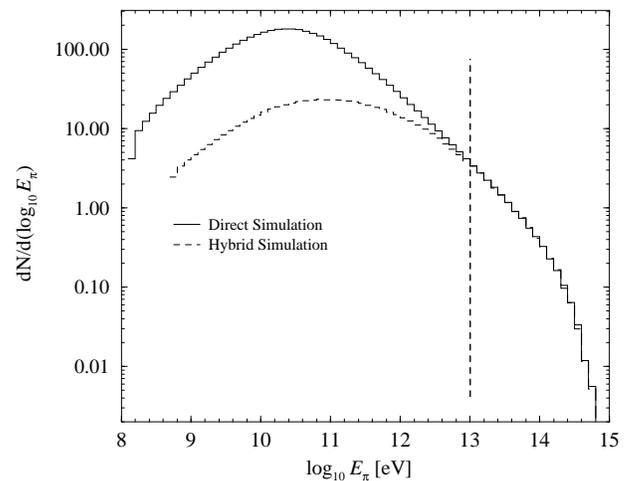}
}
\caption{Energy distribution of pions in showers initiated by primary
protons at $E=10^{15}$ eV. The dashed curve is the energy distribution
of the pions actually treated in our hybrid simulation procedure
using a hybrid energy threshold of $10^{13}$ eV. 
The solid curve shows the energy distribution of pions which are 
explicitly tracked in a direct simulation. Pions which decay
are not shown.}
\label{PionsEnergyPlot}
\end{figure}

In Fig.~\ref{PionsEnergyPlot} we plot the energy distribution of 
pions actually treated in our hybrid simulation procedure (dashed 
\begin{table*}
\caption{Average values of different observables and 
standard deviation of their distributions
obtained by direct and hybrid simulations of 
5,000 vertical pion showers with fixed
interaction point $X_{0}$=5 g/cm$^{2}$, and primary energy 
$E=10^{16}$ eV. The predictions of SIBYLL 1.7, SIBYLL 2.1 and QGSjet98
are presented. The
energy threshold in the hybrid calculation is $0.01~E=10^{14}$ eV.
\label{comparison}} 
\renewcommand{\arraystretch}{1.5}
\begin{tabular}{c|cc|cc|cc} \hline \hline

Model & \multicolumn{2}{c|}{SIBYLL 1.7}&\multicolumn{2}{c|}{SIBYLL 2.1}&\multicolumn{2}{c}{QGSjet98}\\  \hline

 & Direct & Hybrid & Direct & Hybrid & Direct & Hybrid \\ \hline 

 $\langle X_{\rm max}\rangle~{\rm [g/cm^2]}$ & 603 & 602 & 587 & 586 & 574 & 576 \\

 $\sigma~(X_{\rm max})~{\rm [g/cm^2]}$ & 49 & 50 & 51 & 49 & 55 & 56 \\ \hline

 $\langle S_{\rm max}\rangle/E ~{\rm [GeV^{-1}]}$ & 0.75 & 0.76 & 0.75 & 0.75 & 0.75 & 0.75 \\

 $\sigma~(S_{\rm max}/E)~{\rm [GeV^{-1}]}$ & $6.8 \times 10^{-2}$  & $6.8 \times 10^{-2}$ & $6.3 \times 10^{-2}$ & $6.2 \times 10^{-2}$ & $6.5 \times 10^{-2}$ & $6.5 \times 10^{-2}$ \\ \hline

 $\langle N_\mu\rangle~(>0.3$ GeV) & $ 5.39 \times 10^4$ & $5.41 \times 10^4$ & $6.10 \times 10^4$ & $6.13 \times 10^4$ & $6.87 \times 10^4$ & $6.91 \times 10^4$ \\ 

 $\sigma~(N_\mu)$ & $1.79 \times 10^4$  & $1.81 \times 10^4$ & $1.86 \times 10^4$ & $1.87 \times 10^4$ & $2.25 \times 10^4$ & $2.28 \times 10^4$ \\ \hline

 CPU Time [min]$^*$& 935 & 33 & 1091 & 41 & 1398 & 79 \\ \hline \hline 

\end{tabular}

$^*$All CPU times illustrated in this work refer to
a 1 GHz AMD Athlon processor.
\end{table*}
line).  For comparison we also show the energy distribution of pions 
that must be explicitly tracked in a direct simulation (solid line)
at the same energy. The comparison is made for $E=10^{15}$ eV 
proton induced showers and for the nominal energy threshold of $0.01~E$. 
In the hybrid approach only the interactions of pions above 
$0.01~E$ are directly simulated and all lower energy pions 
are replaced with parametrizations.  For a primary energy $10^{15}$ eV 
we typically treat 1 out of 10 pions  of energy $\sim 3\times 10^{10}$ eV 
as can be obtained from the figure. This explains the saving in 
CPU time achieved with the hybrid code with respect to the 
direct simulation -- a factor about 7 for the particular 
energy shown here. This factor rapidly increases with
energy. Already at $E=10^{16}$ eV the hybrid calculation is about 25
times faster than a direct simulation (Table~\ref{comparison}).
For applications which do not depend on the number of muons
this factor increases even further. At $10^{15}$ eV about 25\% of the CPU
time is spent on correctly tracking the numerous muons that are decay
products of charged pions and kaons
which do not initiate hadronic showers,
and hence do not enter the parametrizations. With increasing energy
the number of mesons which decay at an energy above the hybrid threshold
decreases, and the number of muons which have to be simulated
explicitly becomes negligible.

To ensure the consistency of our simulation approach,
we have compared full simulations of pion showers to hybrid 
simulations for the same initial energy and depth
using several energy thresholds. 
We find a very good agreement between the average values of the
different observables and their fluctuations in the direct and
hybrid simulations. Table~\ref{comparison} 
compares the direct simulations and the hybrid method
for 5,000 vertical pion showers with fixed first interaction 
point at $X_{0}$=5 g/cm$^{2}$, energy $10^{16}$ eV, and for the different 
hadronic models. It is very important to note that the differences 
between the two methods of calculation are much smaller than those 
introduced by the different hadronic interaction models, i.e.
by using the hybrid approach 
we do not lose sensitivity to the models we are considering. 

\begin{figure}[h]
\centerline{ 
\includegraphics[width=8.5cm]{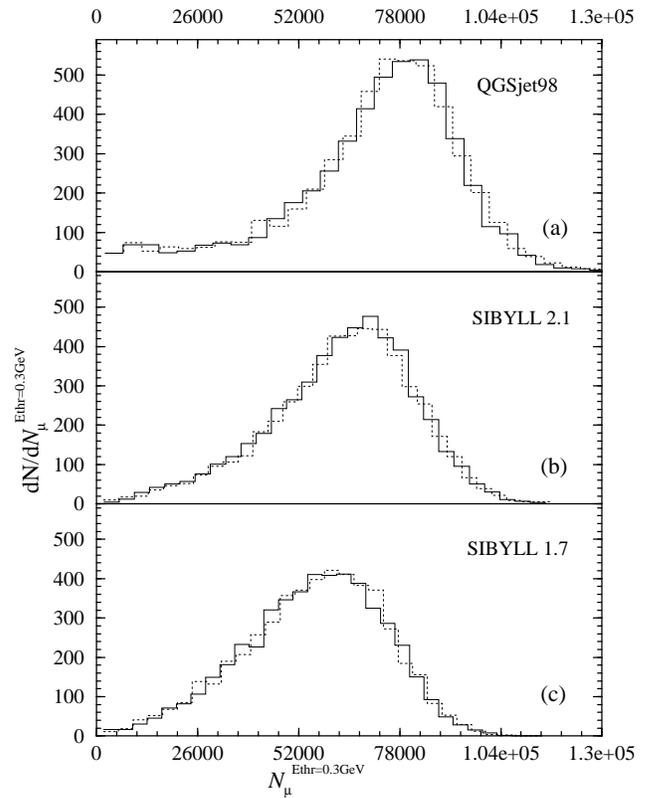} 
}
\caption{Shower distribution in number of muons of energy 
above 0.3 GeV at sea level.
Results are shown for 5,000 vertical pion-initiated showers of energy 
$E=10^{16}$ eV, at fixed interaction point $X_0=5$ g/cm$^{2}$
for different hadronic models. The solid line represents fully 
simulated showers while the dotted line shows hybridly simulated showers 
with meson energy threshold $E/100$.}
\label{Nmu_DistrPion}
\end{figure}
In Fig.~\ref{Nmu_DistrPion} we plot the
distribution of the number of muons with energy above 0.3 GeV
at sea level
for vertical pion-induced showers of energy $10^{16}$ eV.
 We compare the direct simulation to the results of the hybrid
approximation. Panel (a) shows this
 comparison for QGSjet98 and panels (b) and (c) are 
 for SIBYLL 2.1 and SIBYLL 1.7, respectively.
 The relative differences in the average number of muons  
 are less than $0.5\%$ 
 for all hadronic interaction models. 
 The same comparison for showers generated by primary pions with 
 incident zenith angle of 45$^{\circ}$ shows larger differences 
 between direct and hybrid simulations, but they are 
 smaller than $2\%$. 
We believe these relatively small errors come mostly from the 
representation of the intrinsic fluctuations in the shower
development and from the interpolation in energy and atmospheric depth
that the code performs.  
Due to the bootstrap-like calculation of the high-energy part of the
library this error increases slowly with energy.
Comparisons with direct
simulations at $10^{18}$ eV show that
the deviations in the average number
of muons is typically $4\%$. Even for QGSjet98, which predicts the largest
fluctuations at this energy, we obtain a very good description of the
distributions. Their width is reproduced with an
error smaller than 3\%.
The hybrid code always tends to underestimate the
number of muons and its fluctuations.

Our results also show a remarkable stability 
 under changes of the energy threshold, from which we conclude that
 the primary to threshold energy ratio we have used 
($E_{\mathrm {thr}}=E/100$) is sufficient to
 achieve a very good description of the average values and 
 fluctuations of observables in nucleon and pion initiated showers.
Using a threshold for mesons
and for the electromagnetic component fixed to
$E_{\mathrm {thr}}^{\mathrm {em}}=E_{\mathrm {thr}}^{\mathrm {mes}}=E/10$,
we still obtain a good agreement for the average values. However we might
not correctly include some of the extreme fluctuations that are
possible in the early development of the showers.

\section{\label{ResultsProtons}Applications}

In this section we apply the hybrid approach described above
to simulate proton-initiated showers at fixed energy. 
These showers show
most transparently the influence of the hadronic interaction model on
air shower observables. In a forthcoming paper we will calculate
predictions for a realistic cosmic ray spectrum with a mixed cosmic ray 
composition consisting of protons and nuclei \cite{AlvarezMuniz02}.

In the following we consider the hadronic interaction models SIBYLL,
and QGSjet. 
 We have created libraries for the model versions 
SIBYLL 1.7 \cite{Fletcher94}, SIBYLL 2.1 \cite{Engel99,Engel01}
and QGSjet98 \cite{qgsjet}. QGSjet and SIBYLL are sufficiently different
to illustrate various important points of how properties of hadronic
interactions are reflected in shower observables. In addition they
are commonly used for the analysis 
of air shower measurements.
In discussing the models we will focus on QGSjet98 and
SIBYLL 2.1 and show SIBYLL 1.7 predictions only for reference purposes 
because many air shower
data have been already analyzed with this model. 

SIBYLL 2.1 
shows a considerable improvement with respect to version 1.7 in
describing the measurements of hadronic interactions at collider energies.
The important changes in SIBYLL are the implementation
of new parton densities and parton saturation, a new model for
diffraction dissociation, and an energy-dependent soft component
\cite{Engel99}.
Nevertheless, at the highest cosmic ray energies,
its predictions are similar to those of 
SIBYLL 1.7.
On the other hand, QGSjet98 predicts a high-energy
extrapolation which is strikingly different from that of SIBYLL.


\subsection{Hadronic Interaction Models\label{Models}}

QGSjet98 and SIBYLL 2.1 
were shown to describe well collider data up to
the highest energies available so far 
(see for instance \cite{Swordy02df}). However,
already the extrapolation to the unmeasured parts of the phase space is
different. These differences are amplified by going from
proton-proton/antiproton to proton-air collisions.

\begin{figure}[ht]
\centerline{
\includegraphics[width=8.5cm]{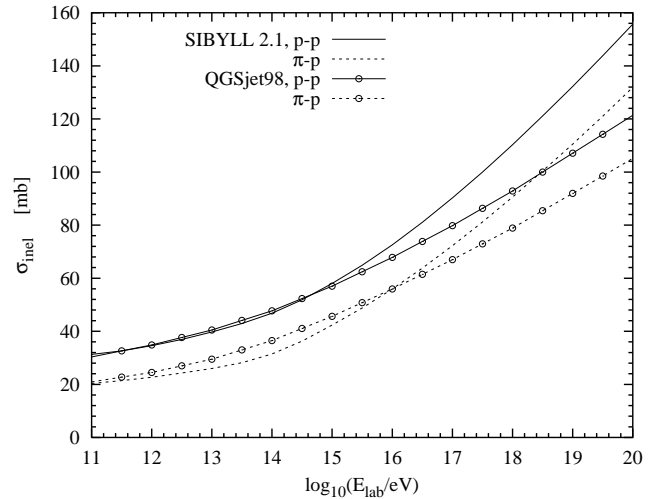}
}
\caption{
Inelastic proton-proton and pion-proton cross sections as predicted by
QGSjet98 and SIBYLL.
\label{fig:model-cs-p}
}
\end{figure}
One of the key features of the hadronic interaction models is their 
prediction on hadron-air cross sections. The proton-air
cross section determines the height of the first interaction in the
atmosphere.
However, it should be emphasized that
the pion- and kaon-air cross sections are also very important for the 
shower development. 
Fig.~\ref{fig:model-cs-p} shows the model cross sections
for proton- and pion-proton collisions which are the input for the
calculation of hadron-air cross sections. In both models free parameters
are adjusted to fit the measured $pp$ and $p\bar p$ cross sections which
cover the energy range from the low end up to 
$E_{\rm lab}\approx 1.7 \times 10^{15}$ eV, 
 i.e. Tevatron center-of-mass energy of $\sqrt{s}=1800$ GeV.
 The model predictions for the pion-proton cross section diverge
 already at much lower energy. The experimental restrictions here
 are much smaller since the pion-proton cross section is experimentally
 known only up to $E_{\rm lab} = 4\times 10^{11}$ eV.
The difference in the high-energy extrapolation of the models arises from
different assumptions on the spatial distribution of partons in protons and
pions. Both models implement the eikonal approximation but differ in
many technical details such as the treatment of inelastic diffraction.
In the following we discuss only the most basic version of the eikonal
model as it is sufficient for explaining the important differences.

In the eikonal model the inelastic cross section is given by 
\begin{equation}
\sigma_{\rm inel} = \int d^2 \vec b \left(1 - \exp\left\{-2\chi_s(s,\vec
b)- 2\chi_h(s,\vec b)\right\}\right)\ , 
\label{eq:inel-cs}
\end{equation}
where the eikonal function is written as the sum of soft and hard
contributions, $\chi_s$ and $\chi_h$. The two-dimensional
impact parameter of the
collision and the squared center-of-mass energy of the collision 
are denoted by $\vec b$ and $s$.
At high energy one has $\chi_h \gg \chi_s$ and the inelastic cross
section is dominated by $\chi_h$, written as
\begin{equation}
\chi_h(s,\vec b) = \frac{1}{2}\sigma_{\rm QCD}(p_\perp^{\rm cutoff},s) 
A(s,b),\hspace*{0.2cm} 
\int d^2 \vec b\;A(s,b) = 1\ .
\end{equation}
The normalized profile function $A(s,b)$ describes the distribution of
partons in the plane transverse to the collision axis. The minijet
cross section $\sigma_{\rm QCD}$ depends on the collision energy and the
transverse momentum cutoff, $p_\perp^{\rm cutoff}$, needed to restrict
the calculation to the perturbative region. For a given energy
dependence of the minijet cross section, only the profile function
$A(s,b)$ determines the inelastic cross section and its energy
dependence. 

Qualitatively, QGSjet is a model which assumes a Gaussian 
profile function \cite{Ostapchenko97a} 
\begin{equation}
A(s,\vec b) = \frac{1}{\pi R^2} \exp\left\{ -\frac{\vec b^2}{
R^2}\right\} ,
\end{equation}
with $R$ being a parameter.
The SIBYLL model is based on the Fourier transform of the electromagnetic 
form factor, assuming that the distribution of gluons in a
hadron is similar to that of the quarks. The corresponding profile
function is energy-independent and is, for example, for proton-proton
scattering \cite{Fletcher94}
\begin{equation}
A (\vec b)  = \frac{\nu^2}{96 \pi}
(\nu|\vec b|)^3 K_3(\nu|\vec b|)\ ,
\label{eq:prof-sib}
\end{equation}
where $K_3$ denotes the modified Bessel function of the third kind and
$\nu \approx 0.7  - 1$ GeV$^{-1}$.

For all $|\vec b| < b_s$ with
$\chi_h(s, b_s) \gg 1$ the saturation limit 
is reached. From
(\ref{eq:inel-cs}) it follows that any further
increase of the minijet cross section would not change the
contribution to the cross section integral from the impact parameter
region $|\vec b| < b_s$.
This allows us to give a rough estimate of the energy dependence of the
inelastic cross section at very high energy. 
For a QCD cross section dependence of $\sigma_{\rm QCD} \sim s^\Delta$,
as is expected within perturbative QCD \cite{Kwiecinski91a},
one gets for a Gaussian profile 
\begin{equation}
b_s^2 \sim R^2 \Delta \ln s 
\end{equation}
and at high energy
\begin{equation}
\sigma_{\rm inel} \approx \int d^2\vec b~\theta(b_s-b) = \pi R^2 \Delta \ln s
\end{equation}
For $R$ being energy-independent the cross section will rise
only logarithmically with the collision energy. However, the parameter
$R$ itself depends on the collision energy through a convolution 
with the parton momentum fractions,  $R^2 \approx R_0^2 +
4 \alpha_{\rm eff}^\prime \ln s$ and 
$\alpha_{\rm eff}^\prime \approx$ 0.11 GeV$^{-2}$.
Hence the QGSjet cross section exhibits a faster than $\ln s$ rise
\begin{equation}
\sigma_{\rm inel} \sim 4 \pi \Delta \alpha_{\rm eff}^\prime \ln^2 s .
\end{equation}
The cross section limit for SIBYLL can be derived in the same way
from Eq.~(\ref{eq:prof-sib})
\begin{equation}
\sigma_{\rm inel} \sim \pi c \frac{\Delta^2}{\nu^2} \ln^2 s\ ,
\end{equation}
where the coefficient $c \approx 2.5$ was found numerically.

Both cross sections satisfy the Froissart bound and exhibit a $\ln^2 s$
energy dependence.
However, the numerical factors are different. Assuming $\Delta \approx
0.25$, then
$4 \pi \alpha_{\rm eff}^\prime \Delta \approx 0.13$ mb and 
$c \pi \Delta^2/\nu^2 \approx 0.2$ mb, 
which explains the faster increase of the
inelastic cross section in the SIBYLL model. A larger power of
$\Delta \approx 0.4$, as implied by 
data from the HERA collider \cite{Abramowicz99a}, even amplifies the
model differences.
However, the difference between the model predictions is smaller than
expected from the arguments above, the reason being a somewhat
smaller minijet cross section assumed in SIBYLL as compared to QGSjet.
The saturation of the parton densities implemented in SIBYLL tames their
rapid growth at small parton momentum \cite{Engel99}.

Information on the profile function can be derived by comparing the
differential elastic cross sections measured at accelerators
with model predictions \cite{Furget90a,Block85}. 
The form factor
approach describes current data reasonably well \cite{Block99a},
whereas a Gaussian profile shows large, systematic deviations and
predicts a wrong curvature.
Although currently available data clearly favor profile
functions derived from electromagnetic form factors, it is not clear
whether this approximation is still good at
ultra-high energy.

\begin{figure}[ht]
\centerline{
\includegraphics[width=8.5cm]{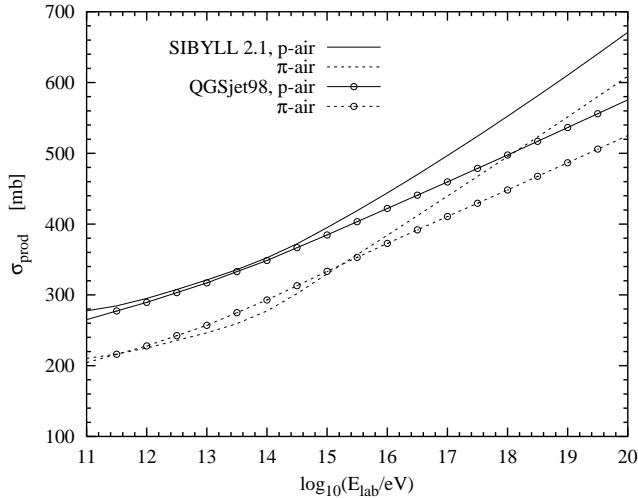}
}
\caption{
Proton- and pion-air production cross section. The production cross
section is defined as the cross section for all collisions in which at
least one new particle is produced. It can be written as $\sigma_{\rm
prod} = \sigma_{\rm tot} - \sigma_{\rm el} -\sigma_{\rm qel}$ where
$\sigma_{\rm tot}$ is the total cross section and $\sigma_{\rm el}$ and
$\sigma_{\rm qel}$ are the elastic and quasi-elastic cross sections
respectively.
\label{fig:model-cs-air}
}
\end{figure}

For hadron-air collisions (Fig.~\ref{fig:model-cs-air}) 
the relative uncertainty in the extrapolated
cross sections is considerably smaller than that of proton-proton and
pion-proton cross sections. The geometrically large size of the
target nucleus (mainly nitrogen or oxygen) dominates the
interaction cross section. At the highest energy considered here the
relative difference is less than 15\%.

\begin{figure}[ht]
\centerline{
\includegraphics[width=8.5cm]{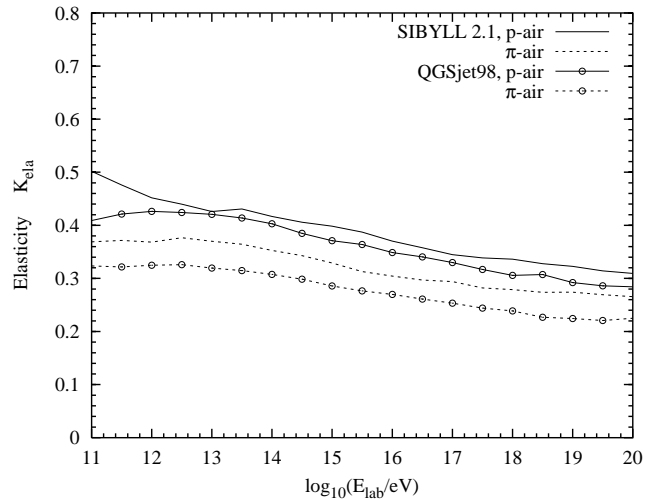}
}
\caption{Mean elasticity in proton-air collisions
as predicted by QGSjet98 and SIBYLL 2.1 (see text).
\label{fig:model-elast}
}
\end{figure}

The evolution of air showers in the atmosphere depends directly on how
much energy is transferred in each hadron interaction into the 
electromagnetic component of the shower. It is common to describe
this energy transfer in terms of the elasticity of the interaction. 
Fig.~\ref{fig:model-elast} shows the mean elasticity of proton- and
pion-air interactions as predicted by QGSjet98 and SIBYLL 2.1. We define the
elasticity of an inelastic interaction (including diffraction
dissociation) as
$ K_{\rm el} = E_{\rm lead}/E_{\rm proj}$ where $E_{\rm lead}$ is the
energy of the most energetic hadron with a long lifetime (i.e. proton,
neutron, $\Lambda$, and charged pions and kaons) and $E_{\rm proj}$ is the
energy of the projectile particle. 
SIBYLL 2.1 consistently predicts more
elastic collisions than QGSjet98 with a relative difference of up to
 17\%.
 Assuming similar other characteristics of hadronic interactions,
 a model with large elasticity predicts air showers which
 develop deeper in the atmosphere.

Other important aspects relevant to air showers 
are the predicted
multiplicity of secondaries and the energy fraction carried by neutral
$\pi^0$'s, which are closely related to the elasticity. 
Neutral pions decay immediately into two photons and feed the
electromagnetic component of the shower. At the highest energies some
neutral pions also interact hadronically because of the enormous time
dilation. On the other hand, the charged particle multiplicity is a
measure of how fast the initial energy is dissipated into many
hadronic low-energy subshowers. It is also a good indicator for the muon
multiplicity since the decaying charged pions are the primary source of
muons.

\begin{figure}[t]
\centerline{
\includegraphics[width=8.5cm]{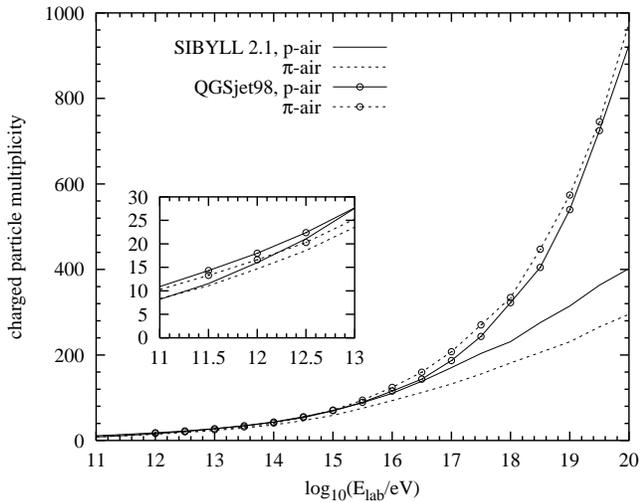}
}
\caption{
Mean multiplicity of charged secondary particles produced
in inelastic proton- and pion-air collisions.
\label{fig:model-ch-mul}
}
\end{figure}

Fig.~\ref{fig:model-ch-mul} shows the mean charged particle multiplicity
in proton- and pion-air collisions as calculated with QGSjet98 and SIBYLL 2.1.
QGSjet98 predicts a power law-like increase of the number of
secondary particles up to the highest energy. In contrast, the SIBYLL
multiplicity exhibits a logarithmic growth similar to $\ln^2 s$ at
high energy. In the energy region from $10^{13}$ to
about $10^{16}$ eV both models predict the same multiplicity in p-air
collisions. However the pion-air multiplicities are significantly
different at all energies.
In SIBYLL different parton densities are used for pions and
protons. The currently implemented parametrizations from Gl\"uck et al.
\cite{Gluck98a,Gluck99a} predict fewer 
partons at low $x$ in pions as compared to protons. The predicted secondary
particle multiplicity is strikingly different at the highest energies.
QGSjet98 predicts more than twice as many secondaries as SIBYLL.
The multiplicity of neutral pions is closely linked to that of
charged particles and hence shows qualitatively the same behavior.

The differences in multiplicity can again be qualitatively understood by
considering Eq.~(\ref{eq:inel-cs}). The minijet cross section predicted
by perturbative QCD describes the inclusive cross section of minijet
pairs. It does not specify how many minijets are produced per single
hadron-hadron collision. The mean minijet multiplicity, $\langle n_{\rm
jet}\rangle$, can only be
calculated after knowing the inelastic cross section
\begin{equation}
\langle n_{\rm jet} \rangle = \sigma_{\rm QCD}/\sigma_{\rm inel}\ .
\end{equation}
The larger multiplicity predicted by QGSjet stems both from the
steeper energy dependence of its minijet cross section and from the
more moderate energy dependence of its inelastic cross section.
A detailed discussion of the relation between the minijet cross
section and secondary particle multiplicity is given in \cite{Engel01}. 

Another difference is emphasized in the inset in
Fig.~\ref{fig:model-ch-mul}. At low energy (i.e. 100 to 1000 GeV lab.
energy) the multiplicity predicted for proton-air collisions is up to
25\% lower in SIBYLL than in QGSjet. Whereas this difference is
unimportant for electromagnetic shower variables,
it becomes observable in the number of low-energy muons produced in the 
decay of charged pions and kaons.

\begin{figure}[ht]
\centerline{
\includegraphics[width=8.5cm]{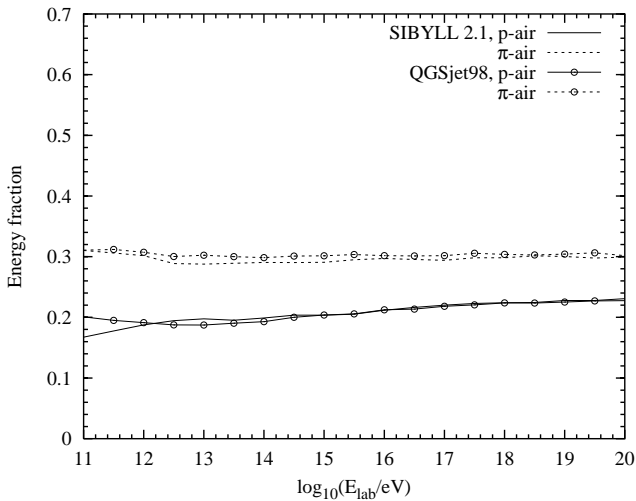}
}
\caption{
Mean energy fraction carried by neutral pions, electrons and photons
in inelastic proton- and pion-air collisions.
\label{fig:model-pi0}
}
\end{figure}
Finally the mean energy fraction carried 
by $\pi^0$'s, $e^\pm$'s and photons is shown in
Fig.~\ref{fig:model-pi0}. Interestingly both models predict 
that the same fraction of
the projectile energy is transferred to the electromagnetic shower
component at high energy. However, the electromagnetic
showers in a SIBYLL 2.1 simulation are more
energetic and less numerous than with QGSjet98.


\subsection{\label{XmaxSmaxResults}Shower Size 
and Depth of Maximum}

$X_{\mathrm {max}}$ and $S_{\mathrm {max}}$ are two  
typical shower parameters measured by fluorescence and Cherenkov
light detectors in several experiments. Knowing the shower energy, the 
mean depth of shower maximum and its fluctuations 
can be used to infer the primary cosmic ray
composition.

\begin{figure}[ht]
\centerline{
\includegraphics[width=8.5cm]{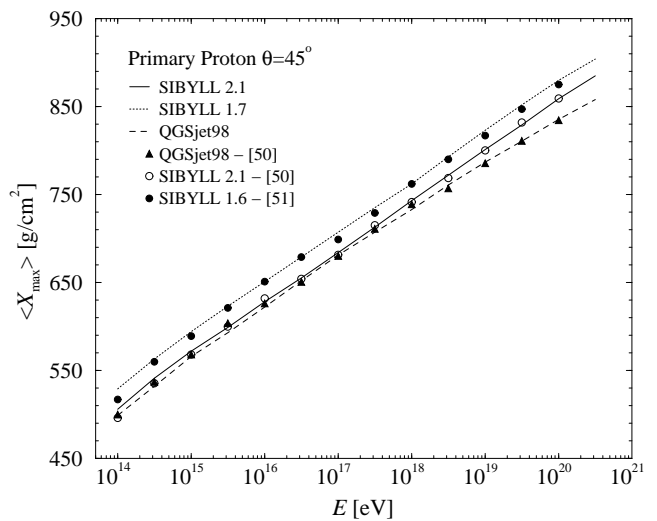}
}
\caption{Average depth of maximum $\langle X_{\rm max} \rangle$ of
proton showers as a function of primary energy. The lines represent
5,000 events generated by our one dimensional method, at $\theta=45^\circ$,
using SIBYLL 1.7 (dotted), SIBYLL 2.1 (solid) and QGSjet98 (dashed).
The symbols show the values of $\langle X_{\rm max} \rangle$ averaged
over 500 showers
obtained with CORSIKA using the thinning procedure.}
\label{EoXmaxPlot}
\end{figure}

Fig.~\ref{EoXmaxPlot} shows the average value 
of $X_{\mathrm {max}}$
as a function of primary energy for proton showers injected at 
a zenith angle $\theta=45^{\circ}$. The lines were produced  
averaging $X_{\mathrm {max}}$ over 5,000 showers.
The predictions of SIBYLL 1.7, SIBYLL 2.1 and QGSjet98 
are shown. The first important feature is that SIBYLL 2.1
predicts smaller $\langle X_{\mathrm {max}} \rangle$ values than
SIBYLL 1.7 by about $22~{\rm g/cm^2}$ from 
$10^{14}$ to $3\times 10^{20}$ eV. The predictions of 
 SIBYLL 2.1 are closer to the values produced by QGSjet98.
 In fact, at energies below about $3\times 10^{17}$ eV the
 difference is smaller than $10~{\rm g/cm^2}$ and it
 increases with energy up to a maximum of $27~{\rm g/cm^2}$
 at $3\times 10^{20}$ eV.
QGSjet98 predicts values of $\langle X_{\mathrm {max}} \rangle$ 
systematically
smaller than the ones produced by both versions of SIBYLL.  
This is due to the much higher average particle multiplicity 
 generated by QGSjet98 and the lower elasticity compared
 to SIBYLL. 
These two features are responsible for the accelerated 
shower development in QGSjet98. 
An interesting
feature is that the proton-proton cross
section in SIBYLL 2.1 is $\sim 25\%$ larger than the one predicted
by QGSjet98 at $10^{20}$ eV, however the larger multiplicity and smaller
elasticity of the latter still dominate producing a smaller  
$\langle X_{\mathrm {max}} \rangle$.

\begin{figure}[ht]
\centerline{
\includegraphics[width=8.5cm]{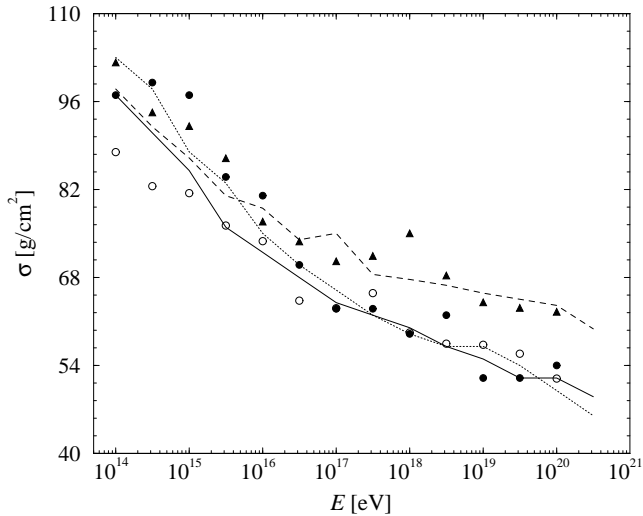}
}
\caption{
Fluctuation of the position of the shower maximum, $\sigma =
\sqrt{
\langle X_{\rm max}^2 \rangle - \langle X_{\rm max} \rangle^2 }$.
The curves have the same meaning as in Fig.~\protect\ref{EoXmaxPlot}.
\label{fig:SigmaXmax}
}
\end{figure}
The width of the $X_{\rm max}$ distribution is a measure of  
the fluctuations of the position of the shower maximum. As shown in
Fig.~\ref{fig:SigmaXmax},  the fluctuations become less important at
very high energy. First of all, the fluctuations due to the position of
the first interaction point are smaller at high energy due to the large
cross section (small mean free path). Secondly, the large multiplicity 
of secondary particles
produces a correspondingly larger number of subshowers. Individual subshowers
will show considerable profile fluctuations as observed at lower energy,
however, due to their large number the total shower profile exhibits 
much smaller fluctuations.

We have verified that the LPM effect \cite{Landau1a,Landau2a,Migdal1a} 
doesn't affect $\langle X_{\rm max} \rangle$ for proton energies 
below $3\times 10^{20}$ eV in agreement with \cite{Cillis99}.
The values of $\langle X_{\mathrm {max}}\rangle$ in  
proton showers at energy $3\times 10^{20}$ and in 
proton showers at the same energy but with the LPM 
artificially ``turned off'' 
are equal within $\sim 1\%$.
\begin{table*}[t]
\renewcommand{\arraystretch}{1.5}
\caption{\label{SmaxXmaxTable0} Mean depth of shower maximum development,
$\langle X_{\mathrm {max}}\rangle$, and shower size 
at depth of maximum $\langle S_{\mathrm {max}} \rangle$, 
in proton-initiated shower with incident zenith
angle $\theta=0^{\circ}$. Each energy represents 5,000 showers simulated
with the hybrid method using SIBYLL 1.7, SIBYLL 2.1, and 
QGSjet98. The width of the corresponding distributions
is given in parenthesis.}
\begin{tabular}{c|cc|cc|cc} \hline\hline
Model & \multicolumn{2}{c|}{SIBYLL 1.7} & \multicolumn{2}{c|}{SIBYLL2.1} & \multicolumn{2}{c}{QGSjet98}\\ \hline
 
 $\lg (E/{\rm eV})$ & $\langle X_{\mathrm {max}} \rangle$ [g/cm$^2]$ & $\langle S_{\mathrm {max}}\rangle/E ~{\rm [GeV^{-1}]}$ & $\langle X_{\mathrm {max}} \rangle {\rm [g/cm^2]}$ & $\langle S_{\mathrm {max}}\rangle/E ~{\rm [GeV^{-1}]}$ & $\langle X_{\mathrm {max}} \rangle {\rm [g/cm^2]}$ & $\langle S_{\mathrm {max}}\rangle /E ~{\rm [GeV^{-1}]}$ \\ \hline
 14.0 & 530 (101) & 0.691 ($1.10 \times 10^{-1}$) & 507 (96) & 0.688 ($1.03 \times 10^{-1}$) & 499 (95) & 0.685 ($9.93 \times 10^{-2}$)\\
 15.0 & 592 (86) & 0.719 ($8.44 \times 10^{-2}$) & 571 (82) & 0.719 ($7.63 \times 10^{-2}$) & 565 (86) & 0.724 ($7.06 \times 10^{-2}$)\\
 16.0 & 647 (72) & 0.735 ($6.07 \times 10^{-2}$) & 626 (71) & 0.734 ($5.57 \times 10^{-2}$) & 625 (78) & 0.736 ($5.25 \times 10^{-2}$)\\
 17.0 & 706 (64) & 0.739 ($4.33 \times 10^{-2}$) & 684 (64) & 0.737 ($4.04 \times 10^{-2}$) & 677 (70) & 0.738 ($3.75 \times 10^{-2}$)\\
 18.0 & 760 (57) & 0.737 ($3.10 \times 10^{-2}$) & 740 (58) & 0.734 ($2.97 \times 10^{-2}$) & 730 (66) & 0.728 ($2.85 \times 10^{-2}$)\\
 19.0 & 822 (55) & 0.723 ($2.80 \times 10^{-2}$) & 799 (55) & 0.718 ($2.56 \times 10^{-2}$) & 785 (66) & 0.708 ($2.36 \times 10^{-2}$)  \\
 20.0 & 878 (51) & 0.694 ($3.46 \times 10^{-2}$) & 856 (51) & 0.690 ($3.43 \times 10^{-2}$) & 832 (62) & 0.683 ($2.85 \times 10^{-2}$)  \\
 20.5 & 901 (46) & 0.671 ($4.29 \times 10^{-2}$) & 880 (47) & 0.666 ($3.98 \times 10^{-2}$) & 853 (56) & 0.662 ($3.55 \times 10^{-2}$)\\ \hline\hline
\end{tabular}
\end{table*}
The large multiplicity of hadronic interactions at energies
above the scale at which the LPM is important, is largely
responsible for this small difference, because it reduces 
the energy of the neutral pions whose decays are the dominant
channel for   
production of high energy photons in the shower. 
Neutral pions then do not produce 
energetic enough photons as to show strong LPM characteristics. 
Even if a high energy neutral pion is created, for instance in
diffractive interactions in which the multiplicity is low, at energies
above $\sim 10^{19}$ eV interactions of neutral pions dominate 
over decay and hence the production of high energy photons
is suppressed \cite{AlvarezMuniz98}. 

Numerical values of $\langle X_{\mathrm {max}} \rangle$ and 
$\langle S_{\mathrm {max}} \rangle$ 
are presented in Table \ref{SmaxXmaxTable0}
for vertical proton induced showers. 
A comparison between  $\langle X_{\mathrm {max}} \rangle$ 
in this table and in Fig.~\ref{EoXmaxPlot} reveals its weak  
dependence on the zenith angle in the 
angular range $\theta=0^\circ-45^\circ$. 
$\langle X_{\mathrm {max}} \rangle$ is fairly insensitive
to changes in atmospheric density profile from 
$\theta=0^\circ$ to $\theta=45^\circ$ and hence it 
is approximately the same when expressed in ${\rm g/cm^2}$.
$\langle S_{\mathrm {max}} \rangle$ also shows a 
weak dependence on the zenith angle and it 
is remarkably independent
of the hadronic interaction models adopted in this work. 
SIBYLL 2.1 produces $\langle S_{\rm max}\rangle$ values smaller 
than those predicted by SIBYLL 1.7 by 1$\%$
in the whole energy range shown in the table.
An interesting aspect about the behavior of 
$S_{\mathrm {max}}/E$ with primary energy is 
that it increases up to energies of $\sim 10^{17}$ eV
and decreases after that 
for all three models. 

\begin{figure}[ht]
\centerline{
\includegraphics[width=8.5cm]{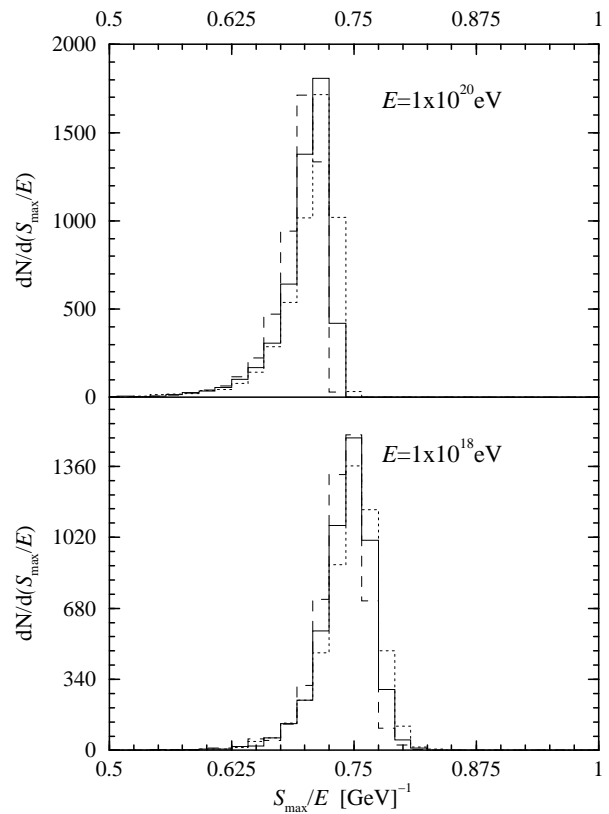}
}
\caption{Distribution of $S_{\mathrm {max}}$ normalized by the
primary energy in GeV.  Results are shown for 5,000 primary proton showers
of energies $10^{18}$ eV (bottom panel) and $10^{20}$ eV (top panel), with
zenith angle $\theta=0^{\circ}$ calculated with the hybrid method using SIBYLL
1.7 (dotted), SIBYLL 2.1 (solid), and QGSjet98 (dashed).}
\label{Smax_DistrProton}
\end{figure}

 In Figs.~\ref{EoXmaxPlot} and \ref{fig:SigmaXmax} we compare 
 our predictions for proton showers
to those obtained in the framework of the CORSIKA code   
using similar (or identical) hadronic interaction models
\cite{Heck98,Heck01,Pryke01}. 
Each of the points generated with CORSIKA in 
Figs.~\ref{EoXmaxPlot} and \ref{fig:SigmaXmax},
represents the mean value of $X_{\rm max}$ over 500 
showers using the thinning procedure. The values of 
$\langle X_{\mathrm {max}} \rangle$ and $\sigma$ calculated by both codes 
for the same models are in very good agreement \cite{Swordy02df}, 
within the larger statistical uncertainty of this particular 
CORSIKA calculation. This 
provides us a further check on the validity of the hybrid simulation 
method.

\begin{figure}[h]
\centerline{
\includegraphics[width=8.5cm]{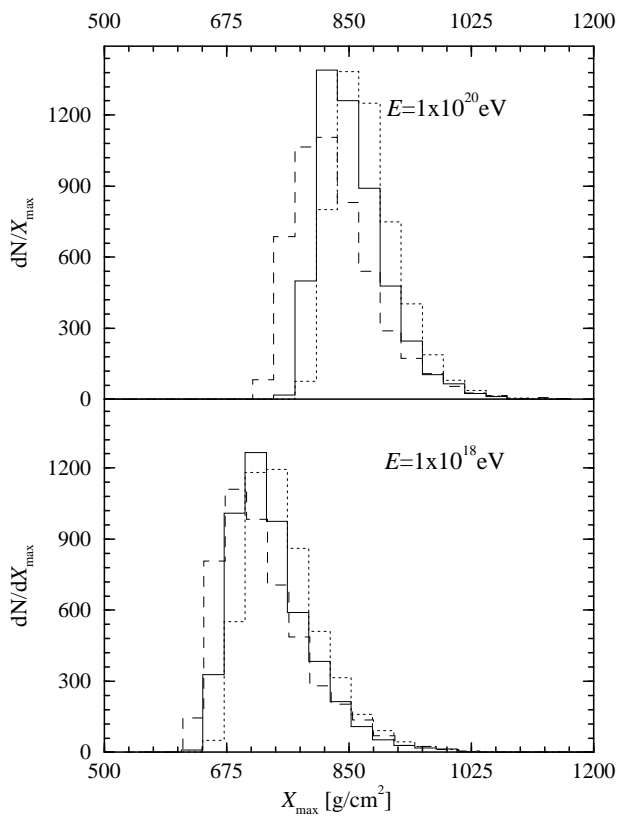}
}
\caption{Distribution of $X_{\mathrm {max}}$.
Results are shown for 5,000 vertical showers generated by primary
protons of energies $10^{18}$ eV (bottom panel) and $10^{20}$ eV
(top panel) calculated with the hybrid method using SIBYLL 1.7 (dotted),
SIBYLL 2.1 (solid), and QGSjet98 (dashed).}
\label{Xmax_DistrProton}
\end{figure}
Fig.~\ref{Smax_DistrProton} shows the distribution of
$S_{\mathrm {max}}$ normalized by the primary energy
in GeV. The top (bottom) histogram represents 5,000 proton-induced
vertical showers at 1 EeV (100 EeV), calculated using
SIBYLL 1.7, SIBYLL 2.1 and QGSjet98. The numerical values of
$\langle S_{\mathrm {max}}\rangle/E$ are shown in Table \ref{SmaxXmaxTable0}.
The distribution of $S_{\mathrm {max}}/E$
is clearly not symmetric around the most likely
value. This is a common feature of the three models and reflects the
asymmetric fluctuations of the various interaction points and
secondary particle multiplicities.

Fig.~\ref{Xmax_DistrProton} shows the distribution of $X_{\mathrm {max}}$,
for the same shower initial parameters as in Fig.~\ref{Smax_DistrProton}. 
The distribution has an asymmetric shape with  
a long tail at large values of $X_{\rm max}$. 
At both energies the tendency of QGSjet98 to predict lower values
of $X_{\rm max}$ is clearly visible. The difference is more 
apparent when compared to the distribution obtained for  
SIBYLL 1.7. The distribution also reflects graphically the
larger fluctuations expected for QGSjet98 compared to SIBYLL.

The fluctuations in $X_{\rm max}$ are directly related
to the relative fraction of diffraction dissociation events generated
in SIBYLL and QGSjet. In particular showers which develop very deep in
the atmosphere are typically those with a diffractive first interaction.
Inelastic diffraction in proton-air collisions
can be subdivided into coherent and incoherent diffraction. The latter
process corresponds to the interaction of the projectile 
with a single nucleon of the target
nucleus and is therefore completely analogous to diffraction in 
proton-proton collisions.
As a multi-channel eikonal model \cite{Kaidalov79}
SIBYLL 2.1 predicts a
growth of the cross section for diffraction dissociation in
proton-proton collisions like $\ln s$
which means that the fraction of low-multiplicity events decreases at
high energy as $1/\ln s$ \cite{Engel99b}. 
In contrast, in QGSjet the fraction of
diffractive events is essentially energy independent (more precisely
proportional to the ratio of the elastic and inelastic cross sections) 
because it is based on the
quasi-eikonal approximation \cite{Boreskov72}. From theoretical grounds
the quasi-eikonal approximation is expected to overestimate 
the diffractive cross section at high energy as it does not 
implement the black disk limit (for a discussion of the black disk 
limit see, for example, \cite{Engel99b}).
On the other hand QGSjet accounts also for coherent diffraction which
is neglected in SIBYLL.


\subsection{Elongation Rate \label{ElongationRate}}

The elongation rate is defined as \cite{Linsley77a,Linsley80a}
\begin{equation}
D_{10} = \frac{d\langle X_{\rm max}\rangle}{d \lg E}.
\end{equation}
It describes the energy-dependence of the position of the shower
maximum. 
The elongation rate reflects changes in the 
cosmic ray composition as well as features of hadronic
interaction at high energy.   Our interest here is in
the relation between elongation rate and 
hadronic interactions.

Most of the charged particles in the shower are electrons and positrons
with energies near the critical energy (81 MeV in air)
from electromagnetic subshowers initiated by photons 
from $\pi^0$-decay.  The mean depth of maximum for
an electromagnetic shower initiated by a photon with
energy $E_\gamma$ is~\cite{Rossi41a}
\begin{equation}
\langle X^{\rm em}_{\rm max}(E_\gamma) \rangle 
= X_0 \ln E_\gamma + C
\end{equation}
where $X_0 \approx 37$g/cm$^2$ is the radiation length in air.
The elongation rate for an electromagnetic shower
is thus $D_{10}^{\rm em} = \ln(10)\times X_0\,\approx\,85$ g/cm$^2$.

A proton-initiated shower consists of a hadronic core
feeding the electromagnetic component primarily through
$\pi^0$ production.  In the approximation of a
hadronic interaction model that obeys Feynman scaling
with energy-independent cross sections,
the energy splitting in the hadronic skeleton of the
shower is independent of energy (i.e. it scales with energy).  
As a consequence, since the
electromagnetic component is dominated by the earliest
(i.e. most energetic) generations of hadronic interactions,
under these assumptions the elongation rate of the hadronic
shower is also $D_{10}^{\rm em}$.  In general, for an incident
nucleus of mass $A$ and total energy $E_0$
(including protons with $A=1$) the depth
of maximum is
\begin{equation}
\langle X_{\rm max}\rangle = X_0\,\ln(E_0/A) + \lambda_A,
\end{equation}
where $\lambda_A$ is the interaction length of
the primary particle.  If the composition
changes with energy, then $\langle A\rangle$ depends
on energy and the elongation rate changes accordingly.

In qualitative analyses of the role of hadronic interactions
in air shower development, an approach analogous to
the treatment of nuclei has often been used.   The depth
of maximum for a proton shower is expressed as
\begin{equation}
\langle X^{\rm had}_{\rm max}(E) \rangle  = \langle X^{\rm em}_{\rm
max}(E/\langle n\rangle) \rangle + \lambda_N\ ,
\label{naive}
\end{equation}
where $\langle n\rangle$ is related to the multiplicity
of secondaries in the high-energy hadronic interactions
in the cascade.  The situation is, however, essentially
more complicated than for a primary nucleus in which
the energy is to a good approximation simply divided into
$A$ equal parts.  In a hadronic cascade
instead there is a hierarchy of energies of secondary
particles in each interaction, and a similar
(approximately geometric) hierarchy of interaction
energies in the cascade.  In this case 
$\langle n\rangle$ has to be understood
as some kind of ``effective'' multiplicity, which does not
have a straightforward definition in general.

The elongation rate derived from Eq.~(\ref{naive}) is
\begin{equation}
\frac{d\langle X^{\rm had}_{\rm max}(E) \rangle}{d \lg E} =
\ln(10) X_0 \left[1 - \frac{d\ln\langle n\rangle}{d\ln E}\right] 
+ \frac{d\lambda_N}{d\lg E},
\label{eq:ER-first}
\end{equation}
which corresponds to the form given by
Linsley and Watson~\cite{Linsley80a},
\begin{equation}
D_{10} = \ln(10) X_0 ( 1 - B_n - B_\lambda ),
\label{eq:ER-orig}
\end{equation}
with 
\begin{equation}
B_n = \frac{d\ln \langle n \rangle}{d\ln E}, \hspace*{0.5cm}
B_\lambda = -\frac{\lambda_N}{X_0} \frac{d\ln \lambda_N}{d \ln E} .
\label{eq:Bn-first}
\end{equation}

For a hadronic interaction model 
with a multiplicity dependence of $\langle n\rangle = n_0
E^\delta$  one gets $B_n = \delta$
{\it provided} all secondaries having the same energy, which is
not the case.

\begin{figure}[ht]
\centerline{
\includegraphics[width=8.5cm]{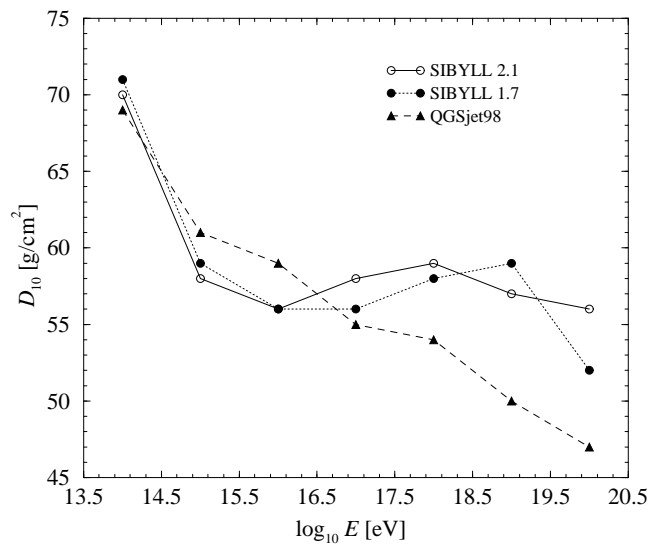}
}
\caption{Elongation rates,
$d\langle X_{\rm max}\rangle/d \lg E$, calculated numerically using showers
simulated with the hybrid method (see text).} 
\label{fig:ElongationRate}
\end{figure}

Because Eq.~(\ref{eq:ER-orig}) is often used to estimate the elongation
rate (see, for example, \cite{Pajares01a}), it is worthwhile to
compare our results with this parametrization.
Fig.~\ref{fig:ElongationRate} shows the elongation rate for
SIBYLL and QGSjet showers as derived from the detailed shower
simulation. All the models show an initial decline from the low-energy
scaling regime as expected.  Then, above $10^{15}$~eV 
the elongation rate for SIBYLL is nearly constant while that
for QGSjet continues to decline.
In addition
SIBYLL 1.7 has a sharp drop of the elongation rate at ultra-high
energy, which we explain below.
In contrast, if we differentiate the curves in 
Figs.~\ref{fig:model-cs-air} and \ref{fig:model-ch-mul}
for cross section and multiplicity and calculate the
elongation rate from (\ref{eq:ER-orig}), using the assumption
of equal sharing of energy among the secondaries, we get completely
misleading results, particularly at low energy.
For example, for QGSjet the predicted elongation rate is about 60
g/cm$^2$ over the entire energy range. The situation is similar for showers 
simulated with SIBYLL 2.1.

The use of the total particle multiplicity for $\langle n \rangle$ 
is not so bad at high energy because
the scaling violation is fully developed in the high energy part of
the shower. However, it is important to note that in Eq.~(\ref{eq:ER-orig}) 
the violation of Feynman scaling and the energy-dependence of the cross
sections are taken into account only for the first interaction. 
All subsequent hadronic interactions are assumed 
to be characterized by Feynman scaling and constant interaction cross
sections. Thus Eq.~(\ref{eq:ER-orig}) is expected to be a good
approximation only in an intermediate energy range around $10^{15} -
10^{16}$ eV.

At higher energy the energy-dependence of the subsequent hadronic
interactions becomes important.
As an illustration, we consider a toy model in which
all final state particles of the first proton interaction in
air are charged pions and have the same energy. At high energy all pions
will interact before decaying. As a first approximation we can write
\begin{equation}
\langle X^{\rm had}_{\rm max}(E) \rangle  = \langle X^{\rm had}_{\rm
max}(E/\langle n\rangle) \rangle + \lambda_N\ ,
\end{equation}
where now the position of maximum of pion induced secondary showers 
is written at r.h.s. Using Eq.~(\ref{eq:ER-first}) to describe the pion
showers one gets for the elongation rate of the entire shower
\begin{eqnarray}
\frac{d\langle X^{\rm had}_{\rm max}(E) \rangle}{d \lg E} = & &
\nonumber\\
& & \hspace*{-2cm}
\ln(10) X_0 \left[1 - \frac{d\ln\langle n(E)\rangle}{d\ln E}
- \frac{d\ln\langle n(E/n(E))\rangle}{d\ln E}
\right]
\nonumber\\
& & \hspace*{-2cm}
+ \frac{d\lambda_N}{d\lg E} + \frac{d\lambda_\pi}{d\lg E}.
\label{eq:ER-second}
\end{eqnarray}
In a model with a power-law increase of the multiplicity with index
$\delta$ this simplifies to
\begin{equation}
\frac{d\langle X^{\rm had}_{\rm max}(E) \rangle}{d \lg E} =
\ln(10) X_0 \left[1 - 2 \delta +\delta^2\right] + \frac{d\lambda_N}{d\lg E} +
\frac{d\lambda_\pi}{d\lg E}.
\end{equation}
Using again the cross sections and multiplicities shown in
Figs.~\ref{fig:model-cs-air} 
and \ref{fig:model-ch-mul} one gets elongation rates
of about 43 and 56 g/cm$^2$ for QGSjet and SIBYLL, respectively.
Given the simplicity of the model the predictions are remarkably
close to the results of the full simulation above $10^{19}$ eV. 
This could be the result
of a cancellation of two effects: on one hand only two successive
hadronic interactions were assumed to be energy-dependent and, on the
other hand, the scaling violation in these interactions was
overestimated by using the total particle multiplicity in
(\ref{eq:ER-second}) and the uniform energy sharing.

Finally it should be mentioned that the sudden drop of the elongation
rate of the showers simulated with SIBYLL 1.7 is due to the onset of
the interaction
of neutral pions. At energies above $10^{19.5}$ eV a substantial number
of $\pi^0$'s does not decay but interacts because of the 
enormous Lorentz dilation.
This effect reduces the mean energy of the particles which feed the 
electromagnetic component of the shower. The change in elongation rate is most
prominent in SIBYLL 1.7 because it generates more
fast (interacting) neutral pions.  


\subsection{\label{MuonResults}Number of Muons}

The number of muons in a shower is an important observable which
depends strongly on the mass of the primary particle and is used
in the studies of the elemental composition of cosmic rays. It also
 directly reflects the hadronic component of the shower and hence
 it is a sensitive probe of the hadronic interactions. 

\begin{figure}
\centerline{
\includegraphics[width=9.0cm]{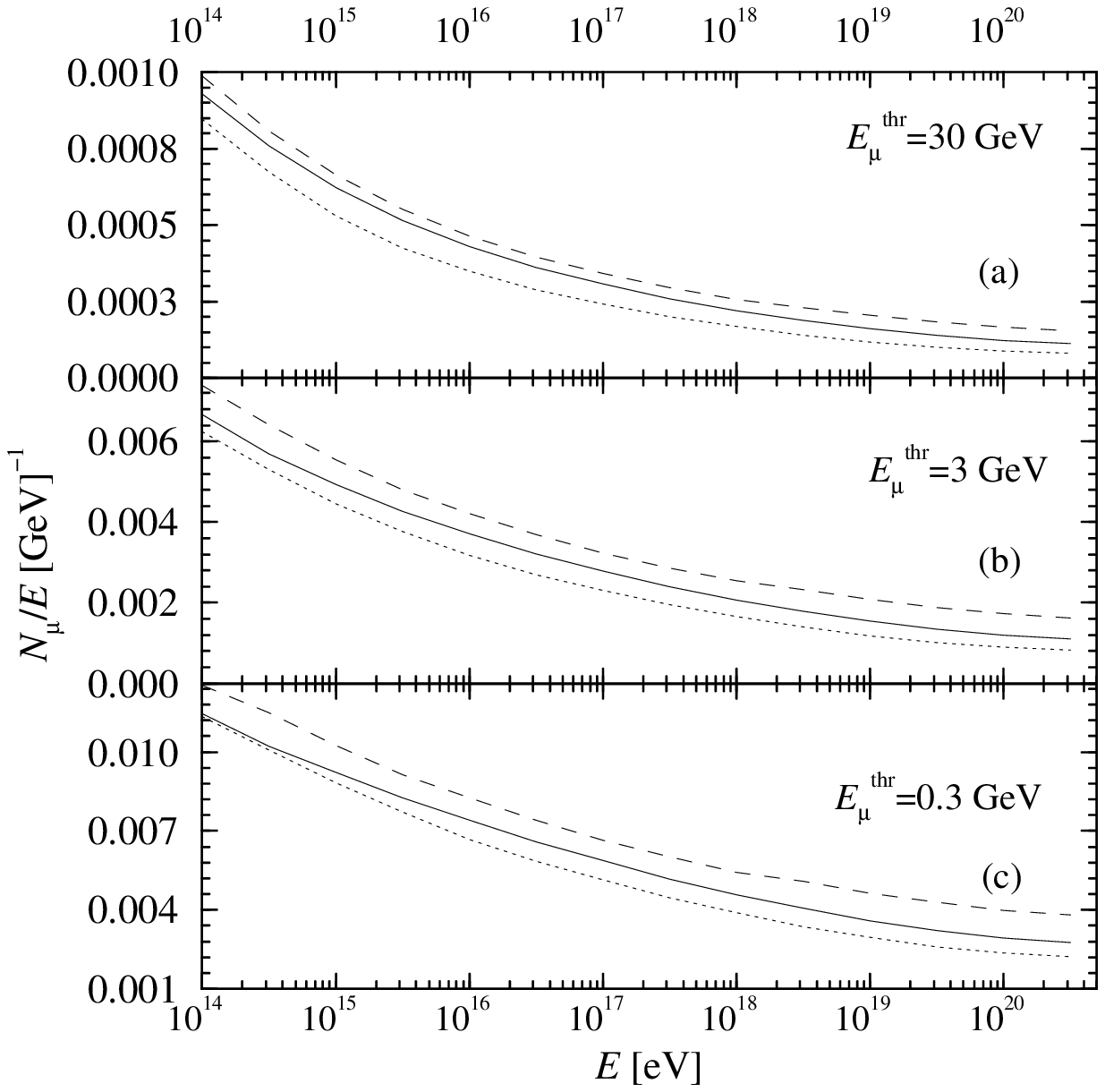}
}
\caption{Average number of muons at sea level $\langle N_\mu\rangle$,
obtained in proton showers with zenith angle $\theta=0^{\circ}$.
Each energy represents 5,000 showers simulated with the hybrid method.
The solid (dotted) line represents the values obtained with
SIBYLL 2.1 (SIBYLL 1.7), while the dashed line illustrates the values for
QGSjet98. Panels (a), (b) and (c) show the average number of
muons with energy above 30 GeV, 3 GeV and 0.3 GeV respectively.}
\label{NmuModelsPlot}
\end{figure}

 We have calculated the average number of muons 
 at sea level ($\langle N_\mu\rangle$)
 with energies above $E_\mu^{\rm thr}=$0.3, 1, 3, 10 and 30 GeV, 
 in proton-initiated showers 
 at zenith angle $\theta=0^\circ$ ($\theta=45^\circ$)
 for the hadronic models SIBYLL 1.7, SIBYLL 2.1, and QGSjet98. 
 Fig.~\ref{NmuModelsPlot} shows
 the energy dependence of the average number of muons normalized
 to the primary energy for the three models. $\langle N_\mu\rangle$ 
 follows approximately a simple power law $(E/E_c)^\alpha$ for energies
 above $\sim 10^{14}$ eV. 

This can be understood on the basis of Heitler's model
\cite{Heitler44} (see also the discussion in \cite{Matthews01})
by assuming that each hadronic interaction
produces in average $\langle n_{\rm tot}\rangle$ secondaries of 
approximately the same
energy. The multiplication of the number of charged pions in a 
shower continues 
until the pions reach a critical energy, $E_c$, at which they are 
assumed to decay. After $N$ generations (i.e. subsequent interactions) 
the energy of the pions reaches the critical energy 
$E_c = E/\langle n_{\rm tot}\rangle^N$.  The number of muons from 
decaying charged pions is thus $N_\mu = \langle n_{\pi^\pm}\rangle^N$.
Eliminating $N$ gives 
\begin{equation}
N_\mu = \left( \frac{E}{E_c}\right)^\alpha,
\hspace*{1cm}\alpha = 
\frac{\ln \langle n_{\pi^\pm}\rangle}{\ln \langle n_{\rm tot}\rangle},
\label{eq:single-power}
\end{equation}
which is the well-known power-law found in data. The index $\alpha$ can be
calculated by using
$\langle n_{\pi^\pm}\rangle \approx \frac{2}{3} \langle n_{\rm tot}\rangle$,
which gives values for $\alpha$ in the range from 0.85 to 0.92. (Assuming
that the charged pion multiplicity is less than 2/3 of the total
multiplicity decreases the values predicted for $\alpha$.)

\begin{table*}
\caption{Parameters $\alpha$ and $E_{\rm c}$ obtained by 
fitting the number of muons in vertical showers at sea level 
using a power law of the form
$N_\mu=(E/E_{\rm c})^\alpha$. The numerical
values of the parameters are presented for the three hadronic
models and for muons with energy above 0.3, 1, 3, 10 and 30 GeV.
\label{tab:fit-naive-0}}
\renewcommand{\arraystretch}{1.5}
\begin{tabular}{c|ccccc|ccccc|ccccc} \hline \hline
 
Model & \multicolumn{5}{c|}{SIBYLL 1.7}&\multicolumn{5}{c|}{SIBYLL 2.1}&\multicolumn{5}{c}{QGSjet98}\\  \hline

~$E_\mu^{\rm thr}$~[GeV]~&~0.3~&~1~&~3~&~10~&~30~&~0.3~&~1~&~3~&~10~&~30~&~0.3~&~1~&~3~&~10~&~30~\\ \hline
 
~$\alpha$~&~0.886~&~0.877~&~0.869~&~0.857~&~0.846~&~0.901~&~0.893~&~0.884~&~0.872~&~0.861~&~0.920~&~0.913~&~0.904~&~0.893~&~0.882~\\ \hline

~$E_{\rm c}$~[GeV]~&~35~&~43~&~67~&~162~&~594~&~39~&~47~&~70~&~161~&~555~&~44~&~53~&~79~&~182~&~638~\\ \hline\hline
 
\end{tabular}

\caption{Parameters $\alpha$ and $E_{\rm c}$ obtained by 
fitting the number of muons in showers with $\theta = 45^\circ$ 
at sea level using a power law of the form
$N_\mu=(E/E_{\rm c})^\alpha$. The numerical
values of the parameters are presented for the three hadronic
models and for muons with energy above 0.3, 1, 3, 10 and 30 GeV.
\label{tab:fit-naive-45}}
\renewcommand{\arraystretch}{1.5}
\begin{tabular}{c|ccccc|ccccc|ccccc} \hline \hline
 
Model & \multicolumn{5}{c|}{SIBYLL 1.7}&\multicolumn{5}{c|}{SIBYLL 2.1}&\multicolumn{5}{c}{QGSjet98}\\  \hline

~$E_\mu^{\rm thr}$~[GeV]~&~0.3~&~1~&~3~&~10~&~30~&~0.3~&~1~&~3~&~10~&~30~&~0.3~&~1~&~3~&~10~&~30~\\ \hline
 
~$\alpha$~&~0.891~&~0.886~&~0.877~&~0.867~&~0.853~&~0.902~&~0.897~&~0.890~&~0.878~&~0.865~&~0.921~&~0.916~&~0.909~&~0.899~&~0.887~\\ \hline

~$E_{\rm c}$~[GeV]~&~65~&~72~&~94~&~195~&~562~&~67~&~74~&~96~&~185~&~523~&~77~&~83~&~107~&~209~&~607~\\ \hline\hline
 
\end{tabular}
\end{table*}
Over the entire energy range from $10^{14}$ eV
to more than $10^{20}$ eV a single power law parametrization can be
used to describe the muon multiplicities for all energy threshold
considered here. In Tabs.~\ref{tab:fit-naive-0} and
\ref{tab:fit-naive-45} we show the corresponding fit parameters for
showers of 0 and 45$^\circ$ zenith angle, respectively. As expected the
critical energy increases with the muon threshold energy. The
energy-dependence of the muon multiplicity is the steepest for low-energy
muons. For a given muon energy threshold, the numerical 
value of $\alpha$ tends to be the highest for the QGSjet98 model.

Already from the simple model discussed above 
it is clear that the power-law index
should be energy-dependent because the multiplicity of the secondary
particles increases with energy. 
Indeed, a careful inspection of the energy 
dependence of $\langle N_\mu \rangle$
shows that the power law index $\alpha$ increases with
the primary energy. However, the observed relative deviation from a 
single power law is always less than 15\%. It is the regularity of this
deviation and the aforementioned physics motivation
which makes it worthwhile to consider the
following, alternative parametrization.

The power-law index is taken to be energy-dependent with
\begin{equation}
\alpha(E) = \left[1+\frac{\ln (3/2)}{\ln\langle
n_{\rm eff}\rangle}\right]^{-1},
\label{eq:index-new}
\end{equation}
where $n_{\rm eff}$ is the geometric average of the charged pion
multiplicity of $N$ successive hadronic interactions.
By construction this effective multiplicity has a 
weak energy dependence, which we approximate by
\begin{equation}
\ln\langle n_{\rm eff}\rangle \approx n_0 + n_1 \ln
\left(\frac{E}{E_0}\right), \hspace*{0.5cm} E_0 = 10^{14}~{\rm eV}.
\label{eq:neff}
\end{equation}
To make the numerical values of $\alpha(E)$ more transparent we 
express the parameters $n_0$ and $n_1$ in terms of 
power-law indices $\alpha_0 = \alpha(E_0)$ and
$\alpha_1 = \alpha(E_1 = 10^{20}{\rm eV})$ 
\begin{eqnarray}
n_0 &=& \frac{\alpha_0}{1-\alpha_0}\ln (3/2)\\
n_1 &=& \frac{\ln (3/2)}{\ln (E_1/E_0)}\left[ \frac{\alpha_1}{1-\alpha_1} -
\frac{\alpha_0}{1-\alpha_0}\right]
\end{eqnarray}
This alternative muon multiplicity parametrization has only three 
free parameters, the
indices $\alpha_0$, $\alpha_1$ and the critical energy $E_c$. 
It gives considerably better fits to the simulation data than the
single power-law parametrization (\ref{eq:single-power}).
\begin{table*}
\caption{Parameters $\alpha_0$, $\alpha_1$ and $E_{\rm c}$ obtained by 
fitting the number of muons in vertical showers at sea level
to Eq.~(\protect\ref{eq:index-new}). The numerical
values of the parameters are presented for the three hadronic
models and for muons with energy above 0.3, 1, 3, 10 and 30 GeV.
\label{tab:fit-new-0}}
\renewcommand{\arraystretch}{1.5}
\begin{tabular}{c|ccccc|ccccc|ccccc} \hline \hline
 
Model & \multicolumn{5}{c|}{SIBYLL 1.7}&\multicolumn{5}{c|}{SIBYLL 2.1}&\multicolumn{5}{c}{QGSjet98}\\  \hline

~$E_\mu^{\rm thr}$~[GeV]~&~0.3~&~1~&~3~&~10~&~30~&~0.3~&~1~&~3~&~10~&~30~&~0.3~&~1~&~3~&~10~&~30~\\ \hline
 
~$\alpha_0$~&~0.858~&~0.838~&~0.819~&~0.780~&~0.745~&~0.887~&~0.870~&~0.850~&~0.820~&~0.787~&~0.855~&~0.834~&~0.809~&~0.775~&~0.736~\\ \hline

~$\alpha_1$~&~0.874~&~0.861~&~0.849~&~0.827~&~0.809~&~0.895~&~0.883~&~0.870~&~0.852~&~0.834~&~0.892~&~0.879~&~0.864~&~0.846~&~0.828~\\ \hline

~$E_{\rm c}$~[GeV]~&~26~&~28~&~39~&~74~&~238~&~33~&~36~&~49~&~97~&~291~&~22~&~23~&~29~&~57~&~179~\\ \hline\hline
 
\end{tabular}
\end{table*}
\begin{table*}
\caption{Parameters $\alpha_0$, $\alpha_1$ and $E_{\rm c}$ obtained by 
fitting the number muons in inclined showers ($\theta=45^\circ$) 
at sea level to Eq.~(\protect\ref{eq:index-new}). The numerical
values of the parameters are presented for the three hadronic
models and for muons with energy above 0.3, 1, 3, 10 and 30 GeV.
\label{tab:fit-new-45}}
\renewcommand{\arraystretch}{1.5}
\begin{tabular}{c|ccccc|ccccc|ccccc} \hline \hline
 
Model & \multicolumn{5}{c|}{SIBYLL 1.7}&\multicolumn{5}{c|}{SIBYLL 2.1}&\multicolumn{5}{c}{QGSjet98}\\  \hline

~$E_\mu^{\rm thr}$~[GeV]~&~0.3~&~1~&~3~&~10~&~30~&~0.3~&~1~&~3~&~10~&~30~&~0.3~&~1~&~3~&~10~&~30~\\ \hline
 
~$\alpha_0$~&~0.873~&~0.863~&~0.842~&~0.819~&~0.764~&~0.892~&~0.883~&~0.867~&~0.839~&~0.802~&~0.858~&~0.848~&~0.828~&~0.794~&~0.751~\\ \hline

~$\alpha_1$~&~0.884~&~0.876~&~0.863~&~0.849~&~0.821~&~0.898~&~0.891~&~0.881~&~0.863~&~0.842~&~0.895~&~0.888~&~0.876~&~0.858~&~0.837~\\ \hline

~$E_{\rm c}$~[GeV]~&~54~&~57~&~66~&~124~&~255~&~61~&~64~&~76~&~128~&~304~&~41~&~42~&~48~&~78~&~189~\\ \hline\hline
 
\end{tabular}
\end{table*}
The numerical values 
obtained by fitting the output of the hybrid simulations are 
shown in Tables~\ref{tab:fit-new-0} and \ref{tab:fit-new-45}.
The relative uncertainties of the parameters $\alpha_0$, $\alpha_1$ 
are about 1\% and 10 to 15\% for $E_c$.

The QGSjet98 model shows the biggest 
change of the power law index from $\alpha_0$ to $\alpha_1$. 
Muon production
in SIBYLL 2.1 is the closest to a simple power law.
The general trend for all three models is that the power law
index decreases with the muon threshold energy.

 The absolute number of muons differs from model to model.
 SIBYLL 2.1 produces more muons than SIBYLL 1.7 but still
 less than QGSjet98 at all energies. The differences between
 the three models increase with energy and reach maximum
 at 10$^{20}$ eV.
 Table~\ref{comp_mu} gives the ratios of $\langle N_\mu\rangle$
 generated by SIBYLL 1.7 and QGSjet98 at sea level to those
 generated by SIBYLL 2.1 in vertical showers at primary energies
 of 10$^{15}$ eV and 10$^{20}$ eV.

\begin{table}
\caption{ Ratios of \protect$\langle N_\mu \rangle$ at sea level
 generated in vertical showers by SIBYLL 1.7 and QGSjet98 to
 those generated by SIBYLL 2.1 (\protect$\equiv$ 1).
\label{comp_mu}}
\renewcommand{\arraystretch}{1.5}
\begin{tabular}{c|ccc|ccc} \hline \hline
$E$~[eV] & \multicolumn{3}{c|}{\protect 10$^{15}$} &
\multicolumn{3}{c}{\protect 10$^{20}$}\\ \hline
$E_\mu^{\rm thr}$~[GeV]~&~0.3~&~3~&~30~&~0.3~&~3~&~30\\ \hline
SIBYLL 1.7 & 0.96 & 0.92 & 0.87 & 0.83 & 0.81 & 0.78 \\
QGSjet98   & 1.11 & 1.12 & 1.06 & 1.37 & 1.41 & 1.35 \\ \hline\hline
\end{tabular}
\end{table}

 It is interesting to observe the dependence of these differences 
 on the muon threshold energy. While for SIBYLL 1.7 the ratio 
 decreases monotonically with the threshold energy, the QGSjet98/SIBYLL 2.1
 ratio shows a more complex behavior. 
The enhanced production of low energy muons in QGSjet98 is related
to the higher charged multiplicity of the model in the 100 - 1000 GeV
range. The differences between the two models decrease for 
$E_\mu^{\rm thr}$ of 30 GeV.

 The number of muons at sea level is sensitive to the incident
 zenith angle. Two competing processes - muon production
 and muon energy loss and decay - determine the dependence on 
 zenith angle. With increasing zenith angle
 both the grammage in which showers develop and the distance
 to the observation level increase. Some additional muons
 are generated in inclined showers due to the larger number 
 of interactions, but also a large fraction
 of the low energy muons (below $\sim 3$ GeV)  
 decay before reaching sea level. Decays win the competition and
 the number of low energy muons decreases with zenith angle.
 At energies above $\sim 10$ GeV, however, most of the muons
 cross the whole atmosphere without decaying, and their number
 at sea level is less sensitive to the injection angle.

\begin{figure}[h]
\centerline{
\includegraphics[width=9.0cm]{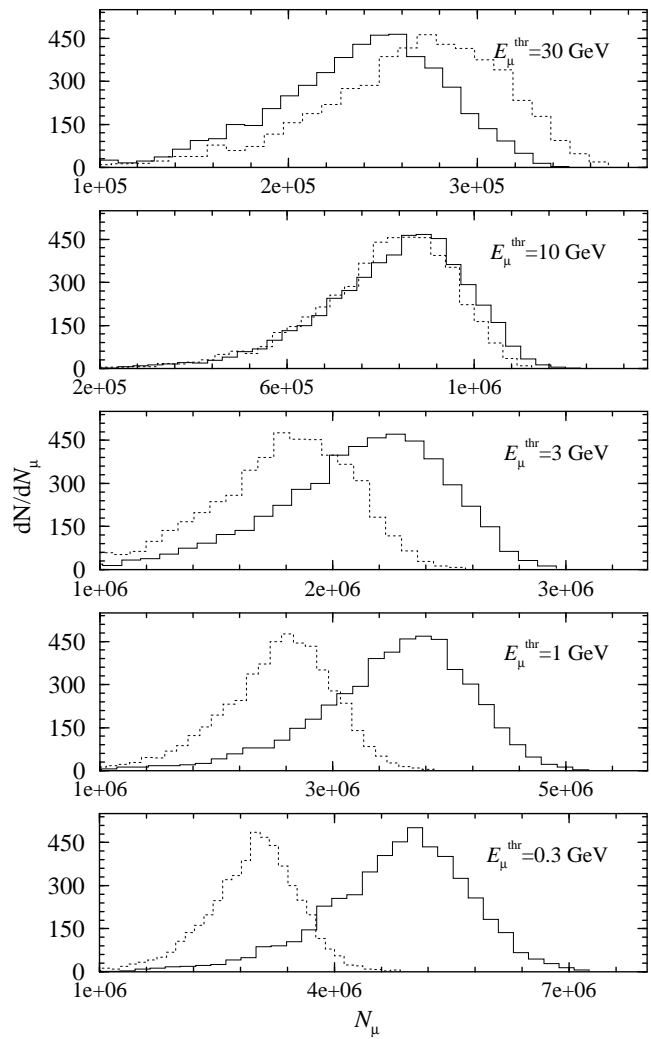}
}
\caption{Shower distribution in number of muons at sea level. The 
results are obtained for 5,000 primary proton showers of energy 
$10^{18}$ eV for different muon energy thresholds. The solid line 
represents vertical showers, while the dotted line illustrates 
showers with zenith angle $\theta=45^\circ$. All showers were 
simulated using SIBYLL 2.1.}
\label{NmuSib21Plot}
\end{figure}

  This behavior is illustrated in Fig.~\ref{NmuSib21Plot} which
 shows the distribution of the number of muons at sea level for
 different $E_\mu^{\rm thr}$ and zenith angles of  
 $0^{\circ}$ and $45^{\circ}$. Each
 histogram represents 5,000 showers initiated by primary 
 protons at 1 EeV using the SIBYLL 2.1 model. 

 At energy above 30 GeV practically all muons cross the atmosphere 
 without decaying. The difference in the number of muons above
 30 GeV between the two zenith angles, depicted in Fig.~\ref{NmuSib21Plot}, 
 is then determined by muon production.
 At large zenith angles shower particles travel for a longer time
 in a more tenuous atmosphere and hence the charged pions have a smaller 
 probability of interaction. As a result more muons are produced
 at $\theta=45^\circ$ than at $\theta=0^\circ$. 

 SIBYLL 2.1 and QGSjet98 predict similar fluctuations in the 
 number of muons. At $E=10^{18}$ eV the width of the 
 shower distribution in muons obtained with QGSjet98 is only
 $\sim 7\%$ larger than in SIBYLL 2.1 for all muon 
 energy thresholds. The difference in the 
 widths at $10^{18}$ eV as obtained with QGSjet98 and SIBYLL 1.7 
 is larger and increases from $\sim 17\%$ at
 $E_\mu^{\rm thr}=0.3$~GeV to $\sim 27\%$ at
 $E_\mu^{\rm thr}=30$~GeV. 

\section{\label{Summary}Summary and outlook} 

We have presented an efficient, one-dimensional hybrid method
to simulate the development of extensive air showers.
 The combination of Monte Carlo techniques for 
 the interactions of the shower particles above a certain
 hybrid energy threshold with a presimulated
 library of pion-induced showers, allows us to simulate 
the development of large statistical samples of air showers up 
to the highest energies observed.  

Previously developed hybrid methods use the average
longitudinal development to describe the numerous subthreshold
showers and are usually limited to the calculation of the total 
number of electromagnetic particles. 
In this paper we have presented a method that accounts for 
fluctuations in the shower development as well as the 
correlations between the different parameters describing
the electromagnetic and muon components of EAS.
By comparing 
 direct simulations with hybrid-simulated showers we have
 determined that the correlation between the hadronic and
 the electromagnetic component is also well reproduced with
 our method. In particular the hybrid method correctly describes
 the correlation between the number of muons and the shower size
at observation level, which is of special relevance to studies
of the cosmic ray composition.  

We have studied the influence of different hadronic interaction
models, namely SIBYLL 1.7, SIBYLL 2.1 and QGSjet98, on shower
observables which are relevant for the determination of the energy
and chemical composition of the primary cosmic ray flux. We 
have presented average values of $X_{\rm max}$, $S_{\rm max}$ and  
the number of muons above 0.3, 1, 3, 10 and 30 GeV at 
sea level, as well as the fluctuations of these quantities.  
The mean muon multiplicities were analyzed with two different models:\\
(i) a simple power-law parametrization,
 which describes the simulation results with a
relative accuracy of better than 10\% (15\% for $E_\mu^{\rm thr}=30$
GeV), and\\
(ii) a model with a slowly changing power-law index, which gives
an excellent description of the data.

 The relation between the features of the interaction models and 
 the shower observables has been extensively discussed. 
 We stress the influence of the different extrapolations of the
 hadronic models to the highest energies on the features of the
 electromagnetic and hadronic component of the shower, and the
 influence of the differences between the models on the number
 of muons predicted by them. Some of these differences    
 exist already at low energies and affect the average numbers 
 of low energy muons.

In QCD-inspired models such as SIBYLL and QGSjet the predictions on
cross sections are inherently linked to the size of Feynman scaling
violation, and hence multiplicity, implemented in the model. 
A model with a steep energy-dependence of the 
hadron-air cross section is usually
characterized by a moderate increase of the multiplicity. Concerning the
position of the shower maximum the effect of large Feynman scaling
violation (or a steeply rising multiplicity) is similar to that of a
steeply rising cross section. This is the reason why the $\langle X_{\rm
max}\rangle$ predictions of
SIBYLL 2.1 and QGSjet98 are rather similar over a wide energy range.
On the other hand, the number of muons at sea level reflects the
multiplicity of low-energy hadrons produced in a shower but depends only
weakly on the hadronic cross sections. Therefore, showers simulated with 
QGSjet produce consistently more low-energy muons than SIBYLL
showers.

As another application of our method we have studied the influence
of the multiplicity, inelasticity, proton-air cross section
on the elongation rate of proton-initiated showers.  We find that the
elongation rate has a complex dependence on the scaling violation and
cross section behaviour of hadronic interaction models. Again, a
steeply rising cross section leads to a decrease of the elongation rate 
qualitatively similar to a steeply rising multiplicity.
Furthermore, a threshold-like behaviour is observed at extremely high
energy. The onset of the hadronic interaction of neutral pions, which
always decay at low energy, leads to a significant decrease of the
elongation rate.

In forthcoming work we will apply our hybrid method to the determination
of the proton-air cross section in experiments that are able
to measure the muon and electromagnetic components at fixed depth, 
as well as in experiments capable of measuring the distribution
of $X_{\rm max}$. Furthermore, we will exploit the fastness of our
method to simulate large statistical samples of showers initiated
by heavy nuclei, with the aim to predict observables that help 
studying the composition of the cosmic ray flux. 

\vspace*{0.5cm}
\noindent
{\bf Acknowledgments}
We are indebted to H.P. Vankov for making his LPM code 
available to us and for many discussions. We thank D. Heck for
providing us with the values of $X_{\rm max}$ for CORSIKA. We also 
acknowledge fruitful discussions with D. Heck,
P. Lipari, S. Ostapchenko, and T. Thouw. 
J.A. Ortiz is supported by CAPES ``Bolsista da CAPES -
Bras\'{\i}lia/Brasil'' and acknowledges Bartol Research
Institute for its hospitality. This research is supported 
in part by NASA Grant NAG5-7009. 
RE, TKG \& TS are also supported by the US Department of Energy contract 
DE-FG02 91ER 40626. TS also acknowledges the hospitality of PCC, 
Coll\`ege de France and a grant from the Minist\`ere de la Recherche 
of France. The simulations presented here were performed on 
Beowulf clusters funded by NSF grant ATM-9977692.


\end{document}